\documentstyle[aps,twocolumn,prb,psfig]{revtex}
\begin{document}
\draft

\wideabs{

\title{Hole concentration induced transformation of \\
the magnetic and orbital structure in Nd$_{1-x}$Sr$_{x}$MnO$_{3}$}
\author{R. Kajimoto and H. Yoshizawa}
\address{Neutron Scattering Laboratory, I. S. S. P., University of
Tokyo, Tokai, Ibaraki, 319-1106, Japan}
\author{H. Kawano\cite{kawano_ad}}
\address{The Institute of Physical and Chemical Research (RIKEN), Wako,
Saitama 351-0198, Japan}
\author{H. Kuwahara\cite{kuwahara_ad}}
\address{Joint Research Center for Atom Technology (JRCAT), Tsukuba,
Ibaraki 305-8562, Japan}
\author{Y. Tokura}
\address{Joint Research Center for Atom Technology (JRCAT), Tsukuba,
Ibaraki 305-8562, Japan  \\ 
and Department of Applied Physics, University of Tokyo, Bunkyo-ku, Tokyo
113-8656, Japan}
\author{K. Ohoyama and M. Ohashi\cite{ohashi_ad}}
\address{Institute for Materials Research, Tohoku University, Sendai
980-77, Japan}

\date{\today}

\maketitle

\begin{abstract}

In order to study the magnetic and crystal structures in the 
Nd$_{1-x}$Sr$_{x}$MnO$_{3}$ system, we have peformed neutron diffraction
measurements on melt-grown polycrystalline samples with $0.49 \le x \le
0.75$. As a function of hole concentration $x$, the system shows
systematic transformation of magnetic and crystal structures, which is
consistently explained by the change of the character of Mn $e_g$
orbitals.  With increasing $x$, the Nd$_{1-x}$Sr$_{x}$MnO$_{3}$ system
exhibits a metallic ferromagnetic state, a metallic A-type AFM state,
and then an insulating C-type AFM state.  The CE-type charge-ordered AFM
state was observed only in the vicinity of $x=1/2$, and it coexists with
the A-type AFM state for $x \gtrsim 1/2$, indicating that the energy
difference between these two states is extremely small.  We also found
that the MnO$_{6}$ octahedra are apically compressed in the CE-type and
A-type AFM states due to the $d(3x^2-r^2)/d(3y^2-r^2)$ or $d(x^2-y^2)$
orbital orderings, respectively, whereas they are apically elongated by
the rod-type $d(3z^2-r^2)$ orbital ordering in the C-type AFM state.  In
addition, a selective broadening of Bragg peaks was observed in the
C-type AFM phase, and its $x$ dependence strongly suggests the charge
ordering for a commsurate value of the hole concentration at either
$x=\frac{3}{4}$ or $x=\frac{4}{5}$.

\end{abstract}

}

\section{Introduction}

The systematic investigation of the phase diagram of the perovskite
manganites was initiated with studies of La$_{1-x}$Ca$_{x}$MnO$_{3}$ in
1950s,\cite{{wollan},{goodenough}} while later in 1980s, a very similar
rich phase diagram was rediscovered for the Pr$_{1-x}$Ca$_{x}$MnO$_{3}$
system.\cite{jirak} A distinct metallic ferromagnetic (FM) state in
these phase diagrams was explained in terms of the double-exchange (DE)
mechanism between the $e_g$ electrons of Mn ions, and the richness of
the phase diagrams were naturally considered as a manifestation of the
strong couplings among the spin, charge, and Jahn-Teller (JT) lattice
distortions.

Recent discovery of the colossal magnetoresistance (CMR) effect in doped
perovskite manganites has renewed interest in these compounds, and
intensive experimental and theoretical efforts have been devoted to
clarify the origin of the CMR effect.  In addition to the DE
interactions as well as the JT distortions, recent studies revealed that
the ordering of the two fold $e_g$ orbitals of Mn ions plays an
essential role to determine physical properties in the hole-doped
manganites. \cite{cheong,maezono,mizokawa,koshibae,kawano,kawano2} For
example, it was recently reported that the underlying $d(x^2-y^2)$-type
orbital ordering leads to a metallic antiferromagnetic (AFM) state
instead of either the metallic FM state or a so-called CE-type
charge/spin ordered insulating state.  In view of newly developed ideas
and of greatly improved experimental techniques, it would be extremely
useful to perform systematic experimental studies concerning a phase
diagram to gain further profound understanding of the interplay between
$e_g$ orbitals and magnetic as well as transport properties in doped
manganites.

For a study of the hole-concentration dependent phase diagram of doped
manganites, we chose the Nd$_{1-x}$Sr$_{x}$MnO$_{3}$ system, because
detailed transport studies have already been performed on this
system. \cite{{kuwahara},{kuwahara2},{kuwahara3}} We have carried out
comprehensive neutron diffraction studies on the
Nd$_{1-x}$Sr$_{x}$MnO$_{3}$ melt-grown polycrystalline samples with the
Sr ion concentration of $x=0.49$, 0.50, 0.51, 0.55, 0.60, 0.63, 0.67,
0.70, and 0.75.  In what follows, we shall demonstrate that the
moderately narrow one-electron bandwidth of the
Nd$_{1-x}$Sr$_{x}$MnO$_{3}$ system yields a variety of physical
properties such as the charge ordering, metal-insulator transition, and
unique magnetic structures as a result of the interplay between the
charge and/or orbital orderings and spin/lattice structures.

As a function of the hole concentration $x$, the
Nd$_{1-x}$Sr$_{x}$MnO$_{3}$ system shows a systematic change of the
magnetic structures.  With increasing $x$, the ground state spin
ordering varies from metallic ferromagnetism to charge ordered CE-type
antiferromagnetism, then to metallic A-type antiferromagnetism, and
finally to insulating C-type antiferromagnetism (F $\to$ CE $\to$ A
$\to$ C) (See Fig. 1).  It can be shown that each spin order corresponds
to its specific orbital order, and the determined crystal structures are
consistent with the corresponding orbital order for indivisual AFM spin
structures.  For example, the structures for the CE-type and A-type AFM
states are characterized by apically compressed MnO$_{6}$ octahedra,
while that of the C-type AFM state consists of apically elongated
octahedra, reflecting their respective layered-type or rod-type orbital
ordering patterns (See Fig. \ref{orbital}).  These systematic changes of
the ordering of the spins and orbitals are well reproduced by the recent
theoretical calculation which takes into account the double degeneracy
of the $e_g$ orbitals. \cite{maezono}

The charge ordering also plays an important role in the
Nd$_{1-x}$Sr$_{x}$MnO$_{3}$ system. \cite{kawano,kuwahara} The CE-type
charge/spin order is formed in the Nd$_{1-x}$Sr$_{x}$MnO$_{3}$ system,
but it is limited in a very narrow region of $x$ around $x=1/2$, and it
coexists with the A-type AFM state in the $x=1/2$ and $x=0.51$ samples.
The coexistence of these two states may be interpreted as the
orbital-order induced phase segregation between the insulating
charge-ordered state and the metallic orbital-ordered state.  In the
C-type AFM phase with $x$ beyond 0.6, the Bragg peaks show a selective
anisotropic broadening, which indicates disorder in the spacing of the
lattice planes along the tetragonal $c$ axis.  We argue that this result
indicates a possible new charge order for a commensurate hole
concentration of either $x = \frac{3}{4}$ or $\frac{4}{5}$.  The
broadening is resulted from both the $d(3z^2-r^2)$-type orbital ordering
and the charge ordering.

The rest of the paper is organized as follows.  The next section briefly
describes the experimental procedures.  The experimental results are
described in Sec. III, where the property of the magnetic and crystal
structures in the Nd$_{1-x}$Sr$_{x}$MnO$_{3}$ system are given with the
results of the Rietveld refinement analysis.  In Secs. IV, the relations
between the magnetic and crystal structures are discussed in detail for
each type of the AFM orderings.  A brief summary is given in Sec. V.

\section{Experimental procedures}

For the present study, the powder samples were prepared by powdering the
melt-grown single crystals, and were pressed into rod shape. The single
crystal samples were grown by floating-zone method. The detailed
procedures of sample preparation were described
elsewhere. \cite{kawano,kuwahara4} The quality of the samples was
checked by X-ray diffraction measurements and inductively coupled plasma
mass spectroscopy (ICP). The results showed that the samples are in
single phase and the hole concentration agrees with a nominal
concentration within 1 \% accuracy.

Neutron diffraction measurements were performed on a powder
diffractometer HERMES and a triple axis spectrometer GPTAS installed in
the JRR-3M research reactor at Japan Atomic Energy Research
Institute. The incident neutron wave lengths of HERMES and GPTAS were
$\lambda = 1.8196$ {\AA} and 2.35 {\AA}, respectively. The collimation
of HERMES is 6$^{\prime}$-open-18$^{\prime}$, while several combinations
of the collimators were utilized at GPTAS, depending on the necessity of
intensity and momentum resolution.  Most of the measurements, especially
those for the structural analysis, were performed on HERMES, but part of
the measurements was performed with GPTAS because stronger intensity of
magnetic reflections are available on this spectrometer due to the high
incident neutron flux.  The samples were mounted in aluminum capsules
with helium gas, and were attached to the cold head of a closed-cycle
helium gas refrigerator. The temperature was controlled within accuracy
of 0.2 degrees. To obtain the structural parameters, the Rietveld
analysis was performed on the powder diffraction data using the analysis
program RIETAN. \cite{izumi}

\section{Magnetic and crystal structures}

\begin{figure}
 \centering \leavevmode
 \psfig{file=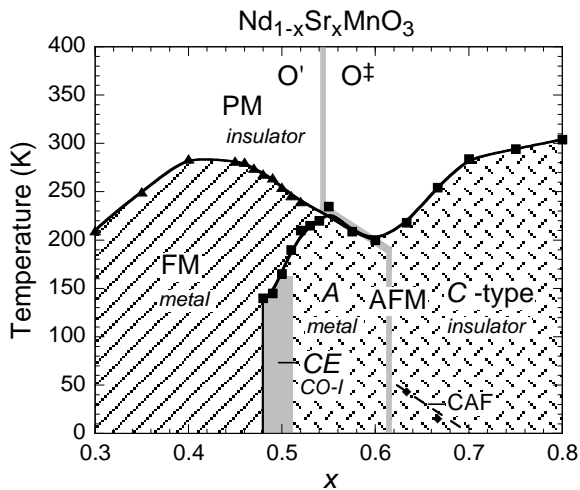,width=0.9\hsize}
 \caption{Phase diagram of
 Nd$_{1-x}$Sr$_{x}$MnO$_{3}$. \protect\cite{kuwahara2,kuwahara3} Each
 phase is denoted by capitalized labels; PM: paramagentic insulating
 phase, FM: ferromangetic order, AFM: antiferromagnetic order, CO-I:
 charge-ordered insulator, CE: CE-type charge/spin order, A: A-type
 antiferromagnetic order, C: C-type antiferromagnetic order, CAF:
 possible canted antiferromagnetic order.}
 \label{phase_diagram}
\end{figure}

We begin with the description of the overall features of the lattice and
magnetic structure of the Nd$_{1-x}$Sr$_{x}$MnO$_{3}$ system by
examining the $x$--$T$ phase diagram for $0.3 \le x \le 0.8$ shown in
Fig. \ref{phase_diagram}. \cite{kuwahara2,kuwahara3} In the distorted
perovskite crystal structure, Mn ions are surrounded by six O ions, and
the MnO$_6$ octahedra form pseudo-cubic lattice, whereas Nd or Sr ions
occupy the body-center position of the pseudo-cubic lattice of the
MnO$_6$ octahedra.  Due to the buckling of the octahedra, however, the
orthorhombic unit cell becomes $\sqrt{2} \times \sqrt{2} \times 2$ of
the cubic cell.  In the concentration region for $0.3 \le x \le 0.8$,
the crystal structure is classified into two phases from the lattice
parameters.  The one is a well-known O$^{\prime}$ phase with $c/\sqrt{2}
< b < a$, \cite{O_prime} and appears in the lower Sr concentration
region for $x \lesssim 0.55$ at room temperature, while the other is a
pseudo-tetragonal O$^{\ddag}$ phase with $a \simeq b < c/\sqrt{2}$ for
$x \gtrsim 0.55$ as indicated in the Fig. \ref{phase_diagram}.  At low
temperatures, on the other hand, the region of the O$^{\prime}$ phase
expands, and the phase boundary shifts towards around $x=0.60$.  In
addition, a monoclinic structure was detected near the low temperature
structural phase boundary near $x \sim 0.60$.  For $0.55 \le x \le
0.60$, a structural transition from the O$^{\prime}$ phase to the
O$^{\ddag}$ phase coincides with the AFM transition temperature $T_{\rm
N}$.

For $x < 0.48$, the ground state is a FM metal.  In the region for $0.50
\lesssim x \lesssim 0.60$, there appears a metallic AFM state with the
layered type AFM ordering, which is called as A-type after
Ref. \onlinecite{wollan}.  With further increasing $x$, the C-type AFM
order was observed in the O$^{\ddag}$ phase.  In this phase, the
resistivity uniformly increases with lowering temperature, and the
sample remains insulating for all temperature, although the temperature
derivative of the resistivity shows an anomaly at $T_{\rm
N}$. \cite{kuwahara3} Only within a small range of the Sr concentration
around $x \sim 0.50$, the system exhibits a charge-ordered insulating
state which is accompanied with the CE-type AFM spin ordering after it
shows the metallic FM state below $T_{\rm C}$ in the intermediate
temperature region.

\subsection{Magnetic structures}

\begin{figure}
 \centering \leavevmode
 \psfig{file=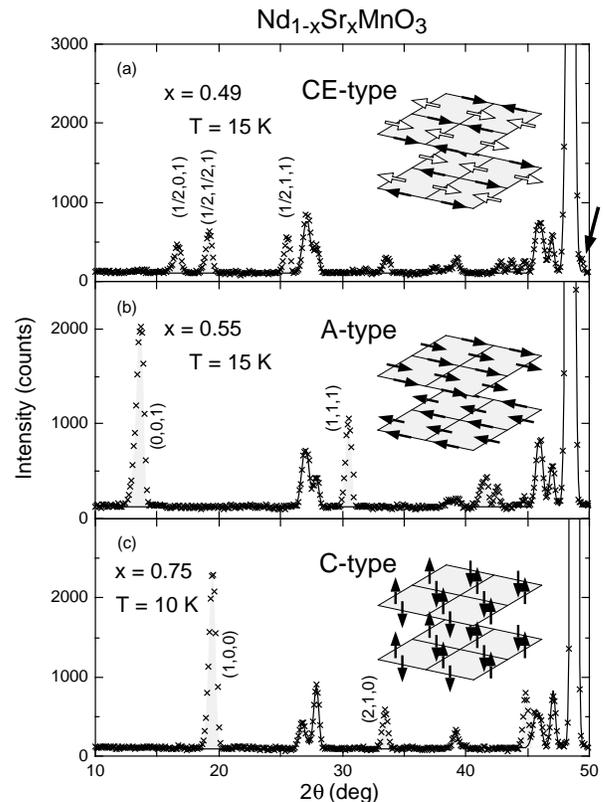,width=0.9\hsize}
 \caption{Low scattering angle portion of the powder diffraction
 patterns for (a) $x=0.49$ at 15 K, (b) $x=0.55$ at 15 K, and (c) $x=
 0.75$ at 10 K which were collected at HERMES.  Solid lines are the
 calculated intensity for the nuclear reflections, and hatched peaks
 represent the AFM Bragg peaks.  Insets are the spin patterns for each
 AFM structures. For the CE-type AFM structure, black and white arrows
 denote the spins of the Mn$^{3+}$ and Mn$^{4+}$ sites, respectively. A
 large arrow in the panel (a) indicates a superlattice peak of the
 lattice distortion due to the CE-type charge ordering.}
 \label{pow_patt}
\end{figure}

The AFM spin ordering yields superlattice reflections in the neutron
diffraction profiles.  Since the positions of the superlattice peaks are
different from each other according to the spin patterns, one can
determine the spin structure from the neutron diffraction profiles.  In
Fig. \ref{pow_patt}, we show the typical powder diffraction patterns
measured at the lowest temperature ($\sim 10$ K) for $x=0.49$, 0.55, and
0.75.  One can clearly recognize the different superlattice reflection
patterns for the CE-type, A-type, and C-type AFM spin arrangements,
respectively.  Cross symbols represent the measured intensity, and solid
lines are the calculated diffraction patterns for the nuclear
reflections obtained by the Rietveld analysis.  The AFM superlattice
reflections are indicated by hatches, and the corresponding spin
ordering patterns are depicted in the insets.

The CE-type spin ordering (Fig. \ref{pow_patt}(a)) is characterized by
the alternate ordering of the Mn$^{3+}$ and Mn$^{4+}$ ions. The spin
ordering pattern in the $ab$ plane is rather complicated, and it stacks
antiferromagnetically along the $c$ axis.  The magnetic reflections for
the Mn$^{3+}$ and Mn$^{4+}$ sublattices are decoupled, and the former
are indexed as $(h/2,~k,~l)$ with $k=\mbox{integer}$ and
$h,\,l=\mbox{odd integer}$, while the latter are indexed as
$(h/2,~k/2,~l)$ with $h,\,k,\,l=\mbox{odd integer}$.

In the A-type spin ordering (Fig. \ref{pow_patt}(b)), the spins order
ferromagnetically in the $ab$ plane with the moments pointing toward the
$a$ axis, and the FM planes stack antiferromagnetically along the $c$
axis. The magnetic reflections appear at $(hkl)$ with $h+k = \mbox{even
integer}$ and $l= \mbox{odd integer}$. As described later, on the other
hand, we observed the monoclinic structure in the A-type AFM phase of
the $x=0.60$ sample, and the FM planes stack along the [1~1~0]
direction, being identical with the case of the monoclinic
Pr$_{1/2}$Sr$_{1/2}$MnO$_{3}$.\cite{kawano}

In the C-type spin ordering (Fig. \ref{pow_patt}(c)), the spins order
ferromagnetically along the $c$ axis, and the neighboring spins in the
$ab$ plane point the opposite direction.  The magnetic reflections are
observed at $(hkl)$ with $h+k = \mbox{odd integer}$ and $l =\mbox{even
integer}$.

We note that, for the $x=0.63$ and 0.67 samples, the FM component was
observed in their magnetization curve below $T_{\rm CA} \sim 45$ K for
the $x = 0.63$ sample or $T_{\rm CA} \sim 15$ K for the $x=0.67$ sample,
\cite{kuwahara3} as indicated by the line CAF in
Fig. \ref{phase_diagram}.  To confirm the existence of the FM component
in these samples, we have measured the temperature dependence of the
(110) and (002) reflections for the $x=0.63$ polycrystalline sample.  If
the FM Bragg scattering appears, the intensity of these reflections
should increase.  Although we have observed a slight increase of the
intensity below $T_{\rm CA}$, its magnitude is no more than the
statistical error.  Even if the FM component exists, it is small to
derive the accurate moment from the powder sample data.

In Table \ref{moment_table}, we summarize the magnetic moments per Mn
site and their directions for all the samples studied.  Interestingly,
we found a clear trend that the direction of the moment is always
parallel to the largest lattice constant (See Table \ref{parm_table}).

\begin{minipage}{\hsize}
 \begin{table}
  \caption{Magnetic structure (MS), magnetic moments, and their
  directions. For the CE-type AFM structure, the magnetic moments are
  shown for both Mn$^{3+}$ and Mn$^{4+}$ sites. The directions of the
  moments for two sites were determined to be identical within
  experimantal accuracy.}
  \label{moment_table}
  \begin{tabular}{crccc}
   $x$ & $T$ (K) & MS & \multicolumn{2}{c}{Moment/Mn ($\mu_{\rm B}$)} \\
   \hline
   0.49 &  15 & CE & 2.9, 2.6 & [100] \\
        &     &  F & 0.8\tablenote{derived from the magnetization curve
   in Ref.\onlinecite{kuwahara2}.} &       \\
        & 160 &  F & 2.6      & [100] \\
   0.50 &  10 & CE & 3.0, 2.7 & [100] \\
   0.51 &  10 &  A & 2.4      & [100] \\
        &     & CE & 1.7, 1.5 & [100] \\
        & 210 &  F & 1.9      & [100] \\
   0.55 &  15 &  A & 3.0      & [100] \\
   0.63 &  10 &  C & 2.8      & [001] \\
   0.67 &  10 &  C & 2.8      & [001] \\
   0.75 &  12 &  C & 2.9      & [001] \\
  \end{tabular}
 \end{table}
\end{minipage}

\subsection{Crystal structures}

\begin{figure}
 \centering \leavevmode
 \psfig{file=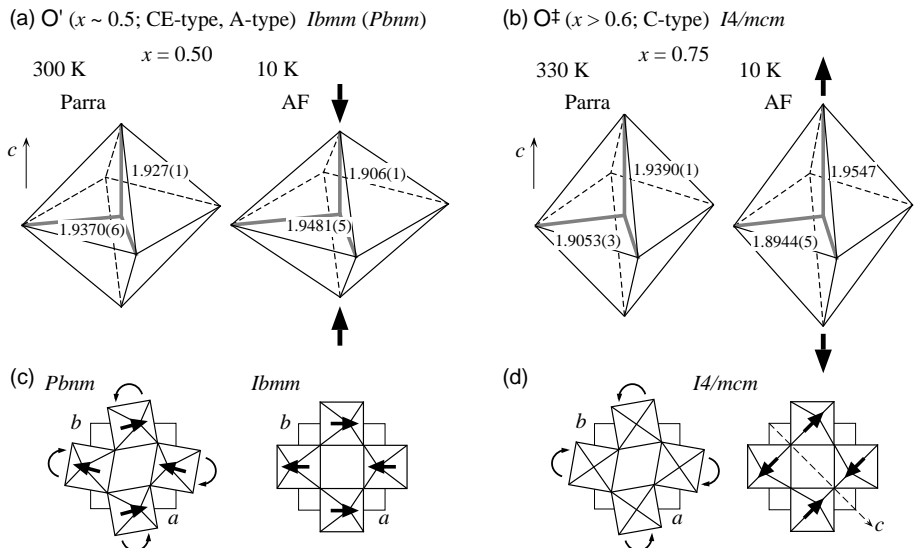,height=7.3cm}
 \caption{Schematic picture of MnO$_6$ octahedra (a) in the O$^{\prime}$
 phase and (b) in the O$^{\ddag}$ phase for the paramagnetic and low
 temperature AFM phases.  (c), (d) The rotation patterns in the basal
 plane for designated structures. For the $I4/mcm$ structure (d), the
 projection of the octahedra onto the one of the [110] planes is also
 depicted in the right panel.}
 \label{octahedra}
\end{figure}

In order to characterize the crystal structures for each phase, we have
performed the Rietveld analysis on neutron powder diffraction patterns
for all the samples observed at selected temperatures.  The obtained
structural parameters are summarized in Table \ref{parm_table}.  The
types of the crystal structure (CS) and of the magnetic structure (MS)
are also listed in the table.  The shapes of MnO$_6$ octahedra for the
O$^{\prime}$ and O$^{\ddag}$ phases are schematically illustrated in
Fig. \ref{octahedra}(a) and (b), respectively.  We examine the
characteristic crystal symmetry for each phase, and discuss the
influence of the distortion of the MnO$_{6}$ octahedra in the following.

In spite of the fact that many orthorhombic perovskite manganites have
the $Pbnm$ ($Pnma$ in another setting) symmetry due to the
GdFeO$_{3}$-type distortion, the measured powder diffraction profiles in
the O$^{\prime}$ phase were well fitted with the orthorhombic space
group $Ibmm$ (or $Imma$).  It should be noted that the $Ibmm$ structure
was also observed in Pr$_{0.65}$Ba$_{0.35}$MnO$_{3}$ at room
temperature,\cite{jirak2} Nd$_{0.5}$Sr$_{0.5}$MnO$_{3}$
(Ref. \onlinecite{caignaert}), and
Pr$_{1/2-x}$Y$_{x}$Sr$_{1/2}$MnO$_{3}$ (Ref. \onlinecite{wolfman}).  To
see the difference of the two structures, the tilting of the MnO$_6$
octahedra for $Pbnm$ and $Ibmm$ are illustrated in
Fig. \ref{octahedra}(c).  In the $Pbnm$ symmetry, the octahedra rotate
both around the $b$ and $c$ axes.  In contrast, the tilting of the
octahedra is restricted only to the $b$ axis in the $Ibmm$ symmetry, and
thereby the $x$ and $y$ coordinates of the inplane oxygen O(2) are fixed
to 1/4.  As a result, two Mn--O bonds in the $ab$ plane have an equal
length.

One can distinguish the $Ibmm$ symmetry from $Pbnm$, in principle,
because $Pbnm$ has a lower symmetry, and there should exist additional
Bragg reflections which are allowed only in the $Pbnm$ symmetry.  In the
paramagnetic phase, however, we observed no additional reflection which
is specific to the $Pbnm$ symmetry, and we tentatively assigned the
space group in the paramagnetic phase as $Ibmm$.  Unfortunately, the
scattering angles of AFM superlattice reflections overlap with the
$Pbnm$ specific reflections in the low temperature AFM phase, and it
prevented us from determining the precise space group in this phase.
Accordingly, we have performed the Rietveld analysis for both space
groups, and obtained almost identical $R$ factors.  For comparison, the
two parameter sets determined for both symmetries on the $x=0.50$ sample
are shown in Table \ref{parm_table}, but only the parameter sets for the
$Ibmm$ symmetry are tabulated for the rest of the samples.

We found that the measured powder diffraction profiles in the
O$^{\ddag}$ phase were well fitted with the tetragonal space group
$I4/mcm$.  Pr$_{0.65}$Ba$_{0.35}$MnO$_{3}$ at 210 K
(Ref. \onlinecite{jirak2}) and La$_{0.5}$Sr$_{0.5}$MnO$_{3}$
(Ref. \onlinecite{sundaresan}) were reported to have the same space
group.  The feature of the space group $I4/mcm$ is also shown in
Fig. \ref{octahedra}(d).  In the $I4/mcm$ symmetry, the MnO$_6$
octahedra rotate only around the $c$ axis, and all the Mn--O bonds in
the $ab$ plane are equal in length.  The octahedra in the $z=1/2$ plane
rotate in the opposite directions to those in the $z=0$ plane.

\twocolumn[\hsize\textwidth\columnwidth\hsize\csname@twocolumnfalse\endcsname
\begin{table}
 \caption{Lattice constants, Mn--O bond lengths, and Mn--O--Mn bond
 angles determined from the Rietveld analysis of the powder profiles.
 CS represents the types of the crystal structure and the meaning of
 symbols O$^{\prime}$ and O$^{\ddag}$ is described in Sec. III.  The
 column MS gives magnetic structures for each samples: symbols P, F, CE,
 A, and C denote the PM, FM, CE-type AFM, A-type AFM, and C-type AFM
 structures, respectively.  $d_{c}$ and $d_{ab}$ denote the Mn--O bond
 lengths along the $c$ axis and within the $ab$ plane.  $\Theta_{c}$ and
 $\Theta_{ab}$ denote the Mn--O--Mn bond angles along the $c$ axis and
 within the $ab$ plane. For $x=0.50$, the parameters obtained by the
 analysis in the $Pbnm$ symmetry are also indicated.  For $x=0.70$ at 10
 K and $x=0.75$ at 10 K, two values for $c/ \protect\sqrt2$ and $d_{c}$
 are shown because the structure parameters were analyzed according to
 the two phase model as described in Sec. VI-A.}
 \label{parm_table}
 \begin{tabular}{crcccccccccc}
  $x$ & $T$ (K) & CS & MS & $a$ (\AA) & $b$ (\AA) & $c/\sqrt{2}$ (\AA)
  & $d_{c}$ (\AA) &  \multicolumn{2}{c}{$d_{ab}$ (\AA)} & $\Theta_c$
  ($^{\circ}$) & $\Theta_{ab}$ ($^{\circ}$) \\ 
  \hline
  0.49 &  15 & O$^{\prime}$ & CE & 5.5129(3) & 5.4409(3) & 5.3198(3) &
  1.907(1) & \multicolumn{2}{c}{1.9491(5)} & 160.9(4) & 166.9(2) \\
       & 160 & O$^{\prime}$ & F & 5.4645(3) & 5.4174(3) & 5.3942(3) &
  1.927(1) & \multicolumn{2}{c}{1.9355(6)} & 163.7(5) & 167.4(3) \\
       & 320 & O$^{\prime}$ & P & 5.4727(3) & 5.4286(3) & 5.3978(3) &
  1.928(1) & \multicolumn{2}{c}{1.9379(6)} & 163.6(5) & 167.9(3) \\
  \hline
  0.50 &  10 & O$^{\prime}$ & CE &  5.5114(4) & 5.4400(4) & 5.3160(4) &
  1.906(1) & \multicolumn{2}{c}{1.9481(5)} & 160.8(4) & 167.2(3) \\
       &     & \multicolumn{2}{c}{in the $Pbnm$ symmetry} & 5.5113(4) &
  5.4400(4) & 5.3160(4) & 1.907(1) & 1.92(2) & 1.98(2) & 160.8(4) &
  167.0(3) \\
       & 300 & O$^{\prime}$ & P & 5.4726(4) & 5.4265(4) & 5.3900(4) &
  1.927(1) & \multicolumn{2}{c}{1.9370(6)} & 163.1(5) & 168.2(3) \\
       &     & \multicolumn{2}{c}{in the $Pbnm$ symmetry} & 5.4726(4) &
  5.4264(4) & 5.3910(4) & 1.927(1) & 1.93(4) & 1.95(4) & 163.1(4) &
  168.1(3) \\
  \hline
  0.51 &  10 & O$^{\prime}$ & A+CE & 5.5075(3) & 5.4441(2) & 5.3131(2)
  & 1.902(1) & \multicolumn{2}{c}{1.9472(4)} & 161.9(4) & 167.7(2) \\
       & 160 & O$^{\prime}$ & A & 5.4999(3) & 5.4446(3) & 5.3242(2) &
  1.905(1) & \multicolumn{2}{c}{1.9453(4)} & 162.3(4) & 168.1(2) \\
       & 210 & O$^{\prime}$ & F & 5.4697(3) & 5.4226(3) & 5.3840(3) &
  1.923(1) & \multicolumn{2}{c}{1.9358(4)} & 163.7(4) & 168.2(3) \\
       & 300 & O$^{\prime}$ & P & 5.4718(3) & 5.4281(3) & 5.3903(3) &
  1.925(1) & \multicolumn{2}{c}{1.9368(5)} & 164.0(5) & 168.4(3) \\ 
  \hline
  0.55 &  15 & O$^{\prime}$ & A & 5.4906(3) & 5.4385(3) & 5.3075(3) &
  1.896(1) & \multicolumn{2}{c}{1.9414(5)} & 163.6(5) & 168.7(3) \\
       & 300 & O$^{\ddag}$ & P & 5.3903(2) & 5.3903(2) & 5.4999(2) &
  1.9445(1) & \multicolumn{2}{c}{1.9218(5)} & 180 & 165.2(2) \\
  \hline
  0.60 &  10 & M & A & 5.365(2) & 5.367(2) & 5.4964(4) & 2.01(3) &
  1.87(6) & 1.90(6) & 180 & 165.0(7) \\
       &     &   &   & \multicolumn{3}{c}{$\beta = 91.225(3)^{\circ}$}
  & 1.88(3) & 1.92(6) & 1.96(6) & & 166.3(7) \\
       & 300 & O$^{\ddag}$ & P & 5.3782(2) & 5.3782(2) & 5.5027(2) &
  1.9455(1) & \multicolumn{2}{c}{1.9162(9)} & 180 & 165.8(4) \\
  \hline
  0.63 &  10 & O$^{\ddag}$ & C & 5.3236(2) & 5.3236(2) & 5.5633(2) &
  1.9669(1) & \multicolumn{2}{c}{1.9009(4)} & 180 & 163.9(2) \\
       &  RT & O$^{\ddag}$ & P & 5.3737(2) & 5.3737(2) & 5.5067(2) &
  1.9469(1) & \multicolumn{2}{c}{1.9139(8)} & 180 & 166.1(4) \\
  \hline
  0.67 &  10 & O$^{\ddag}$ & C & 5.3190(2) & 5.3190(2) & 5.5588(3) &
  1.9653(1) & \multicolumn{2}{c}{1.898(1)} & 180 & 164.4(4) \\
       & 300 & O$^{\ddag}$ & P & 5.3637(3) & 5.3637(3) & 5.5005(3) &
  1.9447(1) & \multicolumn{2}{c}{1.9085(9)} & 180 & 167.1(5) \\
  \hline
  0.70 &  10 & O$^{\ddag}$ & C & 5.3222(3) & 5.3222(3) & 5.5603(4) &
  1.9659(1) & \multicolumn{2}{c}{1.8968(5)} & 180 & 165.5(2) \\
       &     &             &   &           &           & 5.5387(6) &
  1.9582(2) &              &                &     &          \\ 
       &  RT & O$^{\ddag}$ & P & 5.3625(3) & 5.3625(3) & 5.5034(3) &
  1.9457(1) & \multicolumn{2}{c}{1.9060(7)} & 180 & 168.2(4) \\
  \hline
  0.75 &  10 & O$^{\ddag}$ & C & 5.3243(4) & 5.3243(4) & 5.5432(3) &
  1.9598(1) & \multicolumn{2}{c}{1.8944(5)} & 180 & 167.1(3) \\
       &     &             &   &           &           & 5.5144(4) &
  1.9496(2) &               &               &     &          \\
       & 330 & O$^{\ddag}$ & P & 5.3717(2) & 5.3717(2) & 5.4843(3) &
  1.9390(1) & \multicolumn{2}{c}{1.9053(3)} & 180 & 170.8(2) \\
 \end{tabular}
\end{table}

]

\cleardoublepage
\newpage

As illustrated in Figs. \ref{octahedra}(c) and (d), the observed
orthorhombic and tetragonal structures are closely related.  If each
space group is denoted by the tiltings of the MnO$_{6}$ octahedra in the
Glazer's terminology,\cite{glazer} $Pbnm$, $Ibmm$, and $I4/mcm$
symmetries are expressed by $a^{+}b^{-}b^{-}$, $a^{0}b^{-}b^{-}$, and
$a^{0}a^{0}c^{-}$, respectively.  Here the positive and negative signs
denote that the octahedra along the tilt axis are tilted in-phase or
anti-phase, and 0 means no tilt.\cite{glazer} Therefore, they can be
derived from the cubic lattice by introducing successive tiltings of the
MnO$_{6}$ octahedra. Comparing the projection of the octahedra onto the
[001] plane in the orthorhombic phase (Fig. \ref{octahedra}(c)) and that
onto the [110] plane in the tetragonal phase (right part of
Fig. \ref{octahedra}(d)), one can see that, as far as the tilting of the
octahedra is concerned, the tetragonal axis [001] coincides with the
[110] axis in the orthorhombic structure.

\begin{figure}
 \centering \leavevmode
 \psfig{file=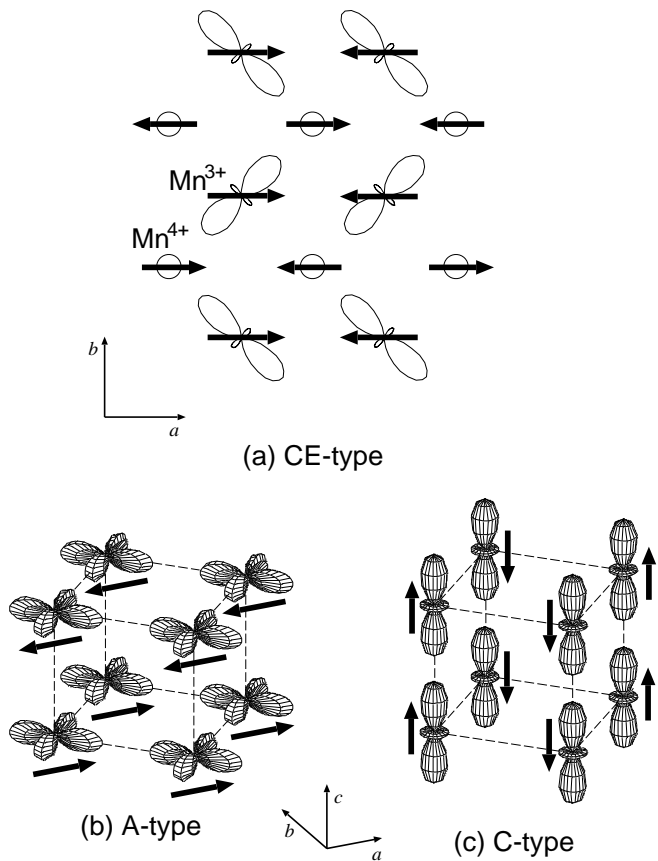,width=\hsize}
 \caption{Schematic picture of the orderings of the $e_g$ orbitals in
 AFM phases. (a) CE-type, (b) A-type, and (c) C-type, respectively. The
 directions of the spins are represented by arrows.}
 \label{orbital}
\end{figure}

In addition to the tilting of octahedra, distortions of the MnO$_{6}$
octahedra provide useful information on the orbital as well as charge
orderings.  As one can see from Table \ref{parm_table}, the two Mn--O
bonds in the $ab$ plane are always longer than those along the $c$ axis
in the O$^{\prime}$ phase.  This feature indicates that the A-type AFM
structure is accompanied with the $d(x^2-y^2)$ type orbital order as
depicted in Fig. \ref{orbital}(b), and this result is supported by the
recent theoretical calculations. \cite{maezono} The $d(x^2-y^2)$ type
orbital order causes a unique magnetic and transport properties due to
the strongly anisotropic couplings within and perpendicular to the
orbital ordered planes.  As we have predicted and demonstrated in the
preceding studies,\cite{kawano,kuwahara2,kuwahara3,mor98a,yoshi} and
will be discussed in Sec. V, this type of orbital order yields an {\it
metallic} A-type AFM state.  It should be noted that one might consider
that the orbital-order-induced anisotropy will be less significant in
the paramagnetic phase, since the difference of the bond length between
the apical and inplane Mn--O bonds becomes rather small at elevated
temperatures.  Surprisingly, however, the anisotropic behavior persists
in high temperature phases.  Very recently, we have demonstrated the
existence of the anomalous anisotropic spin fluctuations in the
paramagnetic and FM phases in
Nd$_{0.50}$Sr$_{0.50}$MnO$_{3}$,\cite{kawano2} indicating that the
$d(x^2-y^2)$ type orbital order has strong influence at high
temperatures.

In the O$^{\ddag}$ phase, on the other hand, the Mn--O bond length along
the $c$ axis is longer than the ones in the $ab$ plane, and this
difference is further enhanced in the AFM phase.  The apically stretched
MnO$_6$ octahedron is consistent with the ordering of the $d(3z^2-r^2)$
orbitals depicted in Fig. \ref{orbital}(c). It is worth to mention that
the recent theoretical calculation confirmed that there appears the
C-type AFM state with the ordering of the $d(3z^2-r^2)$ orbitals in the
higher doping region.\cite{maezono}

Finally, we would like to mention the CE type charge/orbital ordering.
The CE-type ordering is characterized by the alternate ordering of the
Mn$^{3+}$ and Mn$^{4+}$ ions and by the ordering of the
$d(3x^2-r^2)$/$d(3y^2-r^2)$ orbitals on the Mn$^{3+}$ sites in the $ab$
plane as depicted in Fig. \ref{orbital}(a).  This type of orbital
ordering doubles a size of the unit cell along the $b$ axis, and
produces the superlattice reflections at $(h,~k/2,~l)$ with $h =
\mbox{even}$, $k = \mbox{odd}$, and $l = \mbox{integer}$.  Even for the
polycrystalline samples, we could observe the superlattice reflections
in the $x=0.49$ and 0.50 samples at $2\theta = 49^{\circ}$ (indicated by
an arrow in Fig. \ref{pow_patt}(a)) which can be indexed as
$(2\frac{1}{2}2)+(2\frac{3}{2}0)$, as was the case of the
Pr$_{1-x}$Ca$_{x}$MnO$_{3}$ system.\cite{jirak} On the other hand, this
type of charge ordering necessitates to consider two independent Mn
sites for the Mn$^{3+}$ and Mn$^{4+}$ ions. Namely we need to treat two
types of distortions in the MnO$_{6}$ octahedra. Clearly, such an
analysis multiplies the number of parameters in a fitting process, and
would yield less reliable structural parameters.  For this reason, we
omitted to treat the doubling of the unit cell due to the orbital
ordering, and performed the Rietveld analysis on the CE-type samples
assuming only the original $Ibmm/Pbnm$ structure.  Consequently, the
obtained Mn--O bond lengths and Mn--O--Mn angles give the averaged
values for two Mn sites.

\section{Influence of the CE-type ordering}

Near $x \sim \frac{1}{2}$, the doped perovskite manganites are usually
expected to show a so-called CE-type charge/orbital/spin superstructure
depicted in Fig. \ref{orbital}(a).\cite{wollan,goodenough,jirak} For the
Pr$_{1-x}$Ca$_{x}$MnO$_{3}$ system, for instance, the CE-type ordering
is observed over a wide range of $0.3 \lesssim x \leq
0.5$.\cite{jirak,yoshi2,kaji98,cox} As shown in
Fig. \ref{phase_diagram}, however, the CE-type ordering is observed only
in a very limited range near $x \sim 0.50$ in the
Nd$_{1-x}$Sr$_{x}$MnO$_{3}$ system.  Note that, the CE-type ordering was
not observed in the $x=0.55$ sample.\cite{kawano}

A distinct feature of the CE-type ordering in the
Nd$_{1-x}$Sr$_{x}$MnO$_{3}$ system is the coexistence with another spin
ordering.  Because the FM order is taken over by the A-type AFM order
near $x=0.5$ in the Nd$_{1-x}$Sr$_{x}$MnO$_{3}$ system (See the phase
diagram in Fig. \ref{phase_diagram}.), the FM order coexists with the
CE-type order in the $x=0.49$ sample, whereas the A-type AFM order
coexists with the CE-type order in the $x=0.51$ sample.  To illustrate
the situation more specifically, we shall describe the behavior of the
$x=0.49$ and 0.51 samples in detail below.  Concerning the AFM side for
$x> 0.50$, similar results were reported very recently on the same
Nd$_{1-x}$Sr$_{x}$MnO$_{3}$ system with $x=0.52$ and 0.54.\cite{mor98}

\begin{figure}
 \centering \leavevmode
 \psfig{file=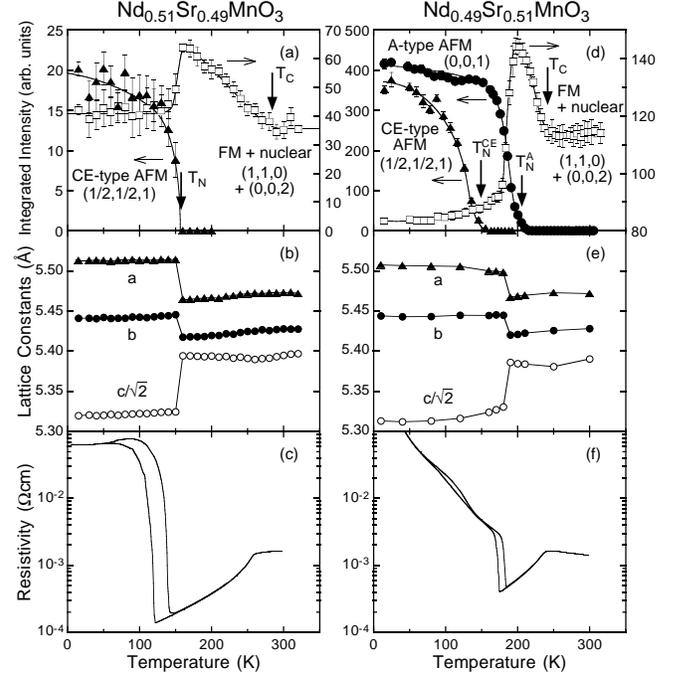,width=\hsize}
 \vspace{2mm}
 \caption{Temperature dependence of the intensities of the magnetic
 Bragg peaks, the lattice constants, and the resistivity for the
 $x=0.49$ FM sample: ((a)--(c)), and for the $x=0.51$ AFM sample:
 ((d)--(f)), respectively. The resisitivity data are reporduced from
 Ref. \protect\onlinecite{kuwahara3}}
 \label{49_51Tdeps}
\end{figure}

Figure \ref{49_51Tdeps}(a)--(f) show the temperature dependences of the
magnetic Bragg peaks, the lattice constants, and the resistivity for the
$x=0.49$ and 0.51 samples.  In the $x=0.49$ sample, the FM spin ordering
was observed below $T_{\rm C} \simeq 280$ K.  As shown in
Fig. \ref{49_51Tdeps}(a), the intensity of the (110) and (002)
reflections increases below $T_{\rm C} \simeq 280$ K due to the FM
order.  Below $T_{\rm N} \simeq 160$ K, it suddenly drops, and the
$(\frac{1}{2}\frac{1}{2}1)$ reflection appears, indicating the formation
of the CE-type AFM order.  It should be noted, however, that the
magnetization study strongly suggests the persistence of the
ferromagntic order below $T_{\rm N}$.  As shown in Table
\ref{moment_table}, the $x=0.49$ sample has the FM moment of $0.8
\mu_{B}$ at 15 K.  Therefore, the FM order coexists with the CE-type AFM
order in the $x=0.49$ sample.

In the $x=0.51$ sample, the behavior of the magnetic ordering was very
similar to that of the $x=0.50$ sample reported in
Refs. \onlinecite{kawano} and \onlinecite{kawano2}.  As shown in
Fig. \ref{49_51Tdeps}(d), the intensity of the $(110)+(002)$ reflections
increases below $T_{\rm C} \simeq 240$ K owing to the onset of the FM
order.  With decreasing temperature, it increases quickly, but shows a
sudden drop at $T_{\rm N}^{\rm A} \simeq 200$ K at which the (001)
A-type AFM reflection appears.  In contrast to the $x=0.49$ FM sample,
the intensity of $(110)+(002)$ reflection in the $x=0.51$ AFM sample has
no magnetic contribution below $T_{\rm N}^{\rm A}$.  The difference
between the intensity above $T_{\rm C}$ and below $T_{\rm N}^{\rm A}$ is
due to the structural transition at $T_{\rm N}^{\rm A}$.  With further
lowering temperature, the $(\frac{1}{2}\frac{1}{2}1)$ CE-type AFM
superlattice reflection appears below $T_{\rm N}^{\rm CE} \simeq 150$ K.
Note that the CE-type ordering suppresses the increase of the intensity
of the (001) A-type AFM reflection below $T_{\rm N}^{\rm CE}$,
indicating that the two spin orderings are strongly correlated.

The temperature dependence of the lattice constants in the $x=0.49$
sample is shown in Fig. \ref{49_51Tdeps}(b).  It shows a weak inflection
at $T_{\rm C}$, while a sharp jump at $T_{\rm N}$, where the $c$ axis
shrinks while the $a$ and $b$ axes expand, being consistent with the
CE-type orbital ordering in the $ab$ plane.  Similarly, the lattice
constants of the $x=0.51$ AFM sample exhibit a large split at $T_{\rm
N}^{\rm A}$ (Fig. \ref{49_51Tdeps}(e)), indicating that the A-type
magnetic transition is accompanied with the structural transition which
stabilizes the $d(x^2-y^2)$-type planar orbital ordering as discussed in
the previous section.  They show, however, no distinct anomaly at
$T_{\rm N}^{\rm CE}$.  In fact, we have carried out detailed structural
analysis on the two powder diffraction data, one is observed at $T=160$
K in the A-type AFM phase for $T_{\rm N}^{\rm CE} < T < T_{\rm N}^{\rm
A}$ and the other at $T=10$ K in the low temperature phase where two
orderings coexist. But we found no difference of crystal structure
between two AFM phases (see Table \ref{parm_table}).  In particular, all
the nuclear reflections in the powder diffraction data at 10 K can be
well fitted to a single structure in spite of the coexistence of the
A-type and CE-type AFM orderings.  These results on the $x=0.51$ sample
demonstrate that the crystal structure of the A-type AFM phase in
Nd$_{1-x}$Sr$_{x}$MnO$_{3}$ is practically indistingushable from that of
the CE-type phase despite the difference of spin structure.\cite{com}

The change of the magnetic structure also has a strong influence on the
behavior of the resistivity.  As shown in Fig. \ref{49_51Tdeps}(c), the
resistivity in the $x=0.49$ FM sample shows the metallic behavior below
$T_{\rm C}$, then a sharp rise at $T_{\rm N}$ due to the CE-type charge
order.  The increase of the resistivity is, however, suppressed below
$\sim$100 K by the FM order with the moments of $0.8 \mu_{B}$ which
coexists with the CE-type AFM spin order.  In the $x=0.51$ AFM sample,
on the other hand, the resistivity shows the metallic behavior below
$T_{\rm C}$, a moderate increase at the onset of the A-type AFM spin
order, and it shows the second increase at $T_{\rm N}^{\rm CE}$ due to
the CE-type charge order as shown in Fig. \ref{49_51Tdeps}(f).  By
comparing this behavior with that of the $x=0.49$ FM sample, the
influence of the spin ordering is clear.  The FM spin order restricts
the resistivity of the $x=0.49$ FM sample at the order of $\rho \sim 5
\times 10^{-2}~\Omega{\rm cm}$ for $T < T_{\rm N}$.  In contrast, the
resistivity of the $x=0.51$ sample exhibits a jump at $T_{\rm N}^{\rm
A}$, it remains an order of $\rho \sim 5 \times 10^{-3}~\Omega{\rm cm}$
in the metallic A-type AFM phase for $T_{\rm N}^{\rm CE} < T < T_{\rm
N}^{\rm A}$, and then increases monotonically below $T < T_{\rm N}^{\rm
CE}$.  It should be noted that the metallic resistivity of the $x=0.51$
AFM sample in the A-type AFM state for $T_{\rm N}^{\rm CE} < T < T_{\rm
N}^{\rm A}$ is of the same order with those of other {\em metallic}
A-type AFM samples.\cite{kawano,kawano2,kuwahara2,kuwahara3,yoshi}

There are several possibilities for the origin of the simultaneous
presence of the CE-type ordering with the FM or A-type AFM spin
orderings in the $x=0.49$ and 0.51 samples.  Scenarios of the
inhomogeneous distribution of the holes in the sample or a canted
magnetic ordering consisting of the CE-type and A-type moments seem to
be consistent with observed results.  The former scenario can be
attributed either to a trivial concentration distribution, or to an
intrinsic phase segregation.  Although it is extremely difficult to
experimentally distinguish these two possibilities, there are some
interesting observations which seem to favor the intrinsic spontaneous
phase segregation in doped manganites near $x \sim \frac{1}{2}$.

As mentioned above, the increase of the CE-type Bragg intensity
suppresses the A-type AFM intensity in the $x=0.51$ sample; in other
words, the CE-type order grows at the expense of the A-type ordered
region.  This fact excludes the possibility of a trivial distribution of
the hole concentration.  In addition, we found that the magnetic moments
for the A-type and CE-type AFM structures lie in the same direction,
i.e., along the $a$ axis (Table \ref{moment_table}).  This result seems
to suggest that a canted magnetic ordering is unlikely in the present
case. \cite{cant_or_not}

We also found that the averaged lattice structure of the CE-type phase
is almost identical with that of the A-type phase in the present
samples.  As shown in the case of the $x=0.51$ sample, the lattice
parameters exhibit little anomaly between two phases.  Since both phases
exhibit the orbital ordering within the basal plane, when the charges
are progressively localized with decreasing $T$, the orbitals are
reorganized with surprisingly small lattice distortions from the
$d(x^2-y^2)$-type orbital order for the A-type AFM ordering to the
$d(3x^2-r^2)$/$d(3y^2-r^2)$-type orbital order for the CE-type ordering.

Combining these observations, we believe that the simultaneous existence
of two states strongly indicates that these two states are very close in
energy, and their relative fraction can be easily varied either by
temperature or by tuning other physical parameters such as one electron
bandwidth, and at the same time this could explain why the CE-type
ordering appears only in a very narrow concentration range of $1 \sim 2$
\% around the $x=1/2$ in the Nd$_{1-x}$Sr$_{x}$MnO$_{3}$ system.  As
discussed in Ref. \onlinecite{kubota}, we argue that this behavior can
be viewed as an effective phase separation between two different orbital
ordered regions which takes place in doped manganites with hole
concentration $x \gtrsim \frac{1}{2}$.  The strong correlation between
the coexistence of the CE-type and A-type orderings and their
resistivity is recently pointed out for doped manganite systems with
$x=\frac{1}{2}$ including two-dimensional single and bilayer systems,
La$_{0.5}$Sr$_{1.5}$MnO$_{4}$ and La$_{1}$Sr$_{2}$Mn$_{3}$O$_{7}$
(Ref. \onlinecite{kubota}).

\section{metallic A-type antiferromagnet}

The most important result in the metallic A-type antiferromagnetic phase
is the fact that all crystal structures in this phase share the common
feature that the lattice spacing in the direction of the AFM stacking is
the smallest (See Fig. \ref{octahedra}).  This salient feature causes an
anisotropy in both magnetic and transport properties, as discussed in
Sec. III-B.  As reported recently, the spin wave dispersion relation in
the metallic A-type AFM Nd$_{0.45}$Sr$_{0.55}$MnO$_3$ exhibits a large
anisotropy of the effective spin stiffness constants between the
intraplanar direction within the FM layers and the interplanar direction
perpendicular to the layers. \cite{kawano,yoshi} It should be noted that
a similar directional anisotropy of the resistivity was also observed in
the A-type AFM samples.\cite{kuwahara2,kuwahara3} These anisotropies in
physical properties are strong evidence of the $d(x^2-y^2)$-type orbital
ordering within the FM layers, and are fully consistent with the
characteristics of the crystal and magnetic structures observed in the
present studies.  In this section, we will focus on the detailed crystal
structures observed in the region of $0.55 \le x < 0.63$ where the
system shows a transition from the paramagnetic O$^{\ddag}$ phase to the
metallic A-type AFM O$^{\prime}$ phase.

\begin{figure}
 \centering \leavevmode
 \psfig{file=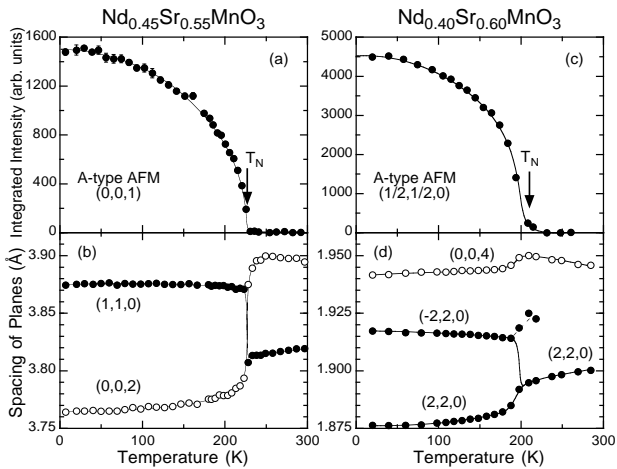,width=0.95\hsize}
 \vspace{2mm}
 \caption{Temperature dependence of the intensities of the AFM Bragg
 peaks and the spacings of the lattice planes. (a),(b) (001) AFM
 Bragg peak and the spacing of (110) and (002) planes for
 $x=0.55$. (c),(d) the $(\frac{1}{2}\frac{1}{2}0)$ AFM Bragg peak and
 the spacing of (220) and (004) planes for $x=0.60$.}
 \label{55_60Tdeps}
\end{figure}

Figures \ref{55_60Tdeps} (a) and (b) show the temperature dependences of
the A-type AFM Bragg peak and the $d$ spacing of the planes of the
(110)/(002) doublet for the $x=0.55$ sample.  The (001) A-type AFM Bragg
peak appears below $T_{\rm N} = 230$ K.  The lattice spacings of the
(002) and (110) nuclear reflections cross at $T_{\rm N}$ due to the
change of the space group from O$^{\ddag}$ to O$^{\prime}$.  The
shrinkage of the $c$ axis in the A-type AFM phase reflects the
$d(x^2-y^2)$-type orbital ordering.

Figures \ref{55_60Tdeps}(c) and (d) show the similar temperature
dependences for the $x=0.60$ sample.  This sample also belongs to the
tetragonal O$^{\ddag}$ phase at the paramagnetic phase, and shows a
first order structural phase transition at $T_{\rm N}$.  In contrast to
the $x=0.55$ sample, however, it has a monoclinic structure whose
unique axis is the $c$ axis in the AFM
phase, as one can clearly see the splitting of the tetragonal (220)
reflection into the monoclinic $(220)+(\bar{2}20)$ reflections below
$T_{\rm N}$. 

We have previously reported that Pr$_{0.50}$Sr$_{0.50}$MnO$_{3}$ also
has the monoclinic structure in the A-type AFM phase, and its crystal
structure belongs to the $P112_1/n$ ($P2_1/c$, cell choice 2) space
group. \cite{kawano} In order to analyze the powder patterns of the
present Nd$_{0.40}$Sr$_{0.60}$MnO$_{3}$ sample collected at 10 K, we
have performed the Rietveld analysis assuming the same $P112_1/n$ space
group at first.  However, we noticed that the $P112_1/n$ space group
predicts too many allowed reflections, compared to the observed Bragg
peaks.  Therefore, in the next step, we fitted the profile with the
space group $I112/m$ ($C2/m$, cell choice 3) which has a higher
symmetry, and we found that the fitting yields the almost equal goodness
with the case of the $P112_1/n$ space group.  We have listed the
parameters obtained with this symmetry in Table \ref{parm_table}.  The
difference of the crystal structure of two space groups are the
following: In both space groups, two unequal Mn sites are placed at
adjacent sites alternately in all directions, but the freedom of the O
sites is much restricted in the case of the $I112/m$ structure.  The
apical oxygen O(1) is placed on the line which connect the nearest Mn
ions along the $c$ axis, and only its $z$ coordinate is allowed to vary,
while the position of the inplane oxygen O(2) and O(3) are confined in
the $ab$ plane.

Unfortunately, the error of the refinement for the $x=0.60$ sample is
the worst among the samples analyzed in the present study ($S = R_{\rm
wp}/R_{\rm e} \simeq 2.2$ for the $x=0.60$ sample, whereas $S$ is less
than 2 for other samples). The main reason for the large $R$ factor may
be that this sample is not in a single phase at 10 K, presumably because
the AFM phase of $x=0.60$ lies just at the boundary between the
orthorhombic O$^{\prime}$ region and the tetragonal O$^{\ddag}$
region.\cite{com2} For the $x=0.63$ sample, we found that $\sim$10 \% of
the sample of the $x=0.63$ sample has the same lattice constants with
those of the $x=0.60$ sample at 10K, and shows the A-type
antiferromagnetism.

This large $R$ factor causes slight ambiguity in identification of the
indices for the closely located peaks such as (004), (220), and (\=220)
in the monoclinic phase, but when the assignment of the three axes were
assumed as labeled in Fig. \ref{55_60Tdeps}(d), the Rietveld analysis
gave the best fit.
 
The $x=0.60$ sample exhibits the same A-type AFM structure with the
$x=0.51$ and 0.55 samples.  However, the monoclinic structure in the
$x=0.60$ sample affects its magnetic structure.  The AFM superlattice
reflections in the $x=0.60$ sample are indexed as ${\bf
Q}=(2n\pm\frac{1}{2},2n'\pm\frac{1}{2},\mbox{even})$ with $n, n' =
\mbox{integer}$, while those in the $x=0.51$ and 0.55 samples are
indexed as $(hkl)$ with $h+k=\mbox{even integer}$ and $l=\mbox{odd}$.
This difference of the reflection conditions indicates that the
propagation vector of the AFM structure for $x=0.60$ is different from
the other A-type samples.  It is rotated by 90$^{\circ}$ from the [001]
axis, and it points towards the [110] direction.  This is consistent
with the fact that the $d$ spacing of (001) remains larger than that of
(110) below $T_{\rm N}$ in this sample.  Such a rotation of the
propagation vector of the A-type AFM ordering was also observed in
another monoclinic sample Pr$_{0.50}$Sr$_{0.50}$MnO$_3$
(Ref. \onlinecite{kawano}).

\section{possible charge order in the C-type AFM insulating phase}

Finally, we shall discuss the features of the insulating C-type AFM
state which appears in the O$^{\ddag}$ phase for $x \geq 0.63$.

\subsection{anomaly in the lattice constant $c$ in the C-type AFM phase}

\begin{figure}
 \centering \leavevmode
 \psfig{file=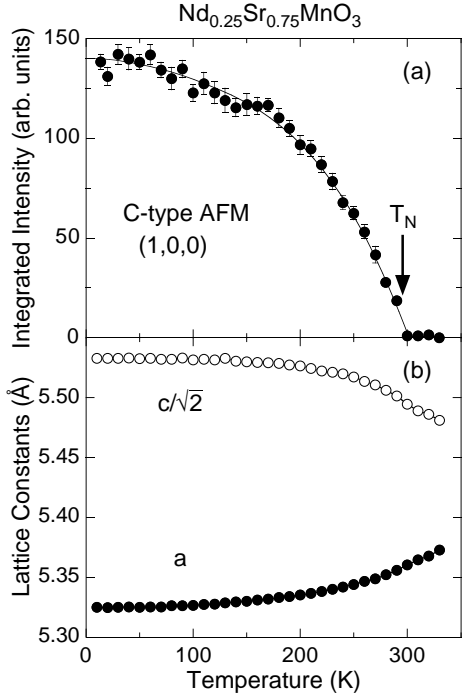,width=0.7\hsize}
 \vspace{2mm}
 \caption{Temperature dependence of the intensities of the AFM Bragg
 peaks and the lattice constants for $x=0.75$.}
 \label{75Tdeps}
\end{figure}

Figure \ref{75Tdeps} shows the temperature dependences of the intensity
for the AFM Bragg peak and of the lattice constants for the $x=0.75$
sample.  The (100) AFM Bragg peak for the C-type spin ordering was
observed below $T_{\rm N} \simeq 300$ K.  The change of the lattice
constants with temperature is very smooth throughout $T_{\rm N}$,
although the difference between the values at 330 K and those at 10 K is
quite large.  With lowering temperature, the length of the $c$ axis
increases whereas the $a$ ($b$) axis decreases.

\begin{figure}
 \centering \leavevmode
 \psfig{file=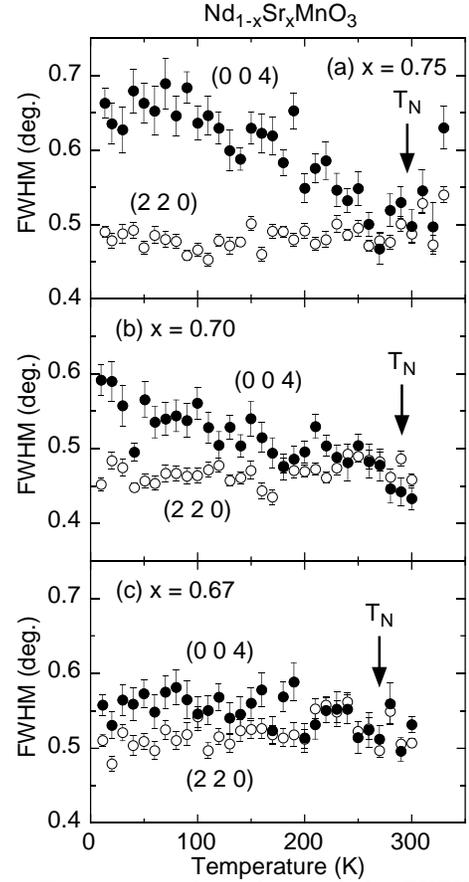,width=0.7\hsize}
 \caption{Temperature dependence of the FWHM of the (004) and (220)
 reflections for (a) $x=0.75$, (b) $x=0.70$, and (c) $x=0.67$.}
 \label{width}
\end{figure}

We would like to stress that we have observed a selective broadening of
the nuclear Bragg reflections.  Figure \ref{width} represents the
temperature dependence of the peak widths (FWHM) of the (004) and (220)
reflections for $x=0.75$, 0.70, and 0.67.  The width of the (220)
reflection remains constant throughout all temperatures.  However, it is
clear that the width of the (004) peak gradually increases below $T \sim
T_{\rm N}$.  Note that the (004) reflection is attributed solely to
nuclear reflection, and no magnetic scattering contributes to this
reflection for the C-type AFM order (see Table \ref{moment_table}).
Comparing the data for three samples depicted in Fig. \ref{width}, one
can see that the $x=0.75$ sample shows the most clear broadening, and it
becomes less distinct as $x$ decreases.

\begin{figure}
 \centering \leavevmode
 \psfig{file=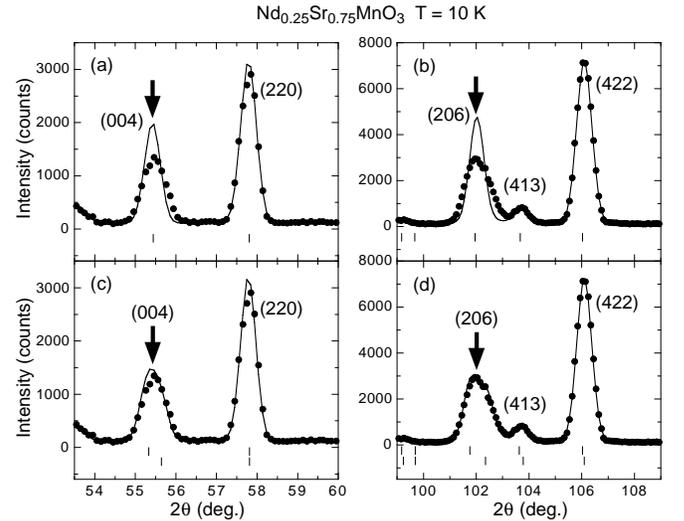,width=\hsize}
 \caption{Portions of the powder diffraction pattern of the $x=0.75$
 sample at 10 K. Solid lines represent the calculated intensity obtained
 by the Rietveld analysis. Calculated peak positions are indicated as
 vertical bars. (a), (b) The refinement was performed with single phase
 of the space group $I4/mcm$. (c), (d) The refinement was performed with
 two phase model described in the text.}
 \label{75prof}
\end{figure}

In order to clarify the origin of the broadening of the nuclear Bragg
reflections, we have examined the powder diffraction patterns and have
found that the broadening was limited to the reflections with the Miller
indices $(hkl)$ of large $l$, for example, (004), (114), (206), (226),
and (008).  In Fig. \ref{75prof}, we show typical examples of the
broadening of the Bragg profiles for the $x=0.75$ sample at 10 K.
Filled circles are the observed intensity profiles, and solid lines are
the calculated intensity obtained by the Rietveld refinement.  One can
clearly see that the widths of the (004) and (206) reflections are wider
than those of (220) and (422).  We first fitted the profile assuming
that the sample is in a single phase with the $I4/mcm$ symmetry, and the
calculated profiles are depicted in Figs. \ref{75prof}(a) and (b).
Despite the fitting is quite good for (220) and (422), the fit to (004)
and (206) is relatively poor.

Because the calculated peak positions are in excellent agreement with
the observed peaks, the symmetry $I4/mcm$ assumed in the analysis cannot
be too far from the true crystal symmetry.  Therefore, one of the
possible reasons for these broadening could be the lowering of the
symmetry to the orthorhombic or monoclinic one, and the resultant
splitting of the original reflections which may be unresolved due to the
moderate angular resolution of the neutron powder diffractometer.  This
possibility, however, will be easily discarded because the peak
broadening is also observed at (00$l$) reflections which will split in
neither the orthorhombic nor monoclinic structure.

Considering the fact that the peak broadening occurs selectively at the
reflections with large $l$, that is, the reflections from the lattice
planes which are nearly perpendicular to the $c$ axis, it is very likely
that it is originated from an anisotropic strain in the system, and a
possible microscopic picture of such a strain is a distribution of the
lattice constant of the $c$ axis.  To ascertain this idea, we have
assumed a simple structural model that the sample consists of two phases
in which they have two different lattice constants for the $c$ axis.  By
keeping other parameters identical for both phases, we have fitted the
observed diffraction pattern to this model, and have obtained
substantially improved results as depicted in Figs. \ref{75prof}(c) and
(d).

\begin{figure}[htbp]
 \centering \leavevmode
 \psfig{file=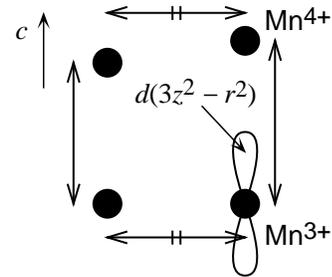,width=0.5\hsize}
 \caption{Schematic picture about the relation of $e_g$ orbital ordering
 and the lattice spacing in the O$^{\ddag}$ phase.}
 \label{Corbital}
\end{figure}

From these facts, we concluded that the observed selective broadening of
Bragg peaks results from the anisotropic strain caused by a distribution
of the $d(3z^2-r^2)$ orbitals.  As stated above, the $e_g$ electrons
occupy the $d(3z^2-r^2)$ orbitals in the O$^{\ddag}$ phase.  When the
charges are localized in the insulating phase, only Mn$^{3+}$ sites have
the $d(3z^2-r^2)$ orbital, and Mn$^{4+}$sites have no $e_g$ electrons.
In Fig. \ref{Corbital}, we illustrated an arrangement of Mn$^{3+}$ with
the $d(3z^2-r^2)$ orbital and Mn$^{4+}$ with no $e_g$ orbital.  Because
the $d(3z^2-r^2)$ orbital extends toward the $c$ direction, the distance
between Mn$^{3+}$ and Mn$^{4+}$ are elongated along the $c$ direction,
whereas in the $ab$ plane the Mn$^{3+}$--Mn$^{4+}$ distance is almost
equal to the Mn$^{4+}$--Mn$^{4+}$ distance.  At high temperatures, the
charges are mobile by thermal activation, which averages out the local
distortion of the lattice spacing along the $c$ axis.  At low
temperatures, on the other hand, the thermal energy is insufficient for
the charges to hop, and the local ordering of the $e_g$ electrons with
the $d(3z^2-r^2)$ orbital may be formed, and it leads to the anomaly in
the lattice constant.  It should be noted that the broadening of the
peak starts below $T_N$ where the C-type AFM spin ordering is formed as
shown in Fig. \ref{width}, and at the same time a temperature derivative
of the resistivity shows an anomaly.\cite{kuwahara3}

\subsection{possibility of the $x=0.8$ charge ordering in the C-type AFM
 phase}

As we described in the previous subsection, we found that the broadening
of the nuclear Bragg peaks becomes clearer as $x$ increases, indicating
that the charge localization/ordering is also progressively stabilized
with the increase of $x$.  Furthermore, the anomaly of the temperature
derivative of the resistivity at $T_{\rm N}$ develops as $x$ increases,
and it is most clearly observed at $x=0.80$.  In addition, the
resistivity of the $x=0.80$ sample itself shows a steep increase at
$T_{\rm N}$ with decreasing temperature. \cite{kuwahara3} Judging from
these facts, it is very likely that the charge ordering associated with
the ${\rm Mn}^{3+}:{\rm Mn}^{4+}$ ratio of $1:4$ is formed below $T_{\rm
N}$.  On the other hand, Jir\'ak {\it et al.} was reported that
superlattice reflections which might be originated from the charge
ordering of ${\rm Mn}^{3+}:{\rm Mn}^{4+} = 1:3$ were observed in their
Pr$_{0.2}$Ca$_{0.8}$MnO$_{3}$ sample whose valence distribution of the
Mn ions was determined to be Mn$^{3+}_{0.25}$Mn$^{4+}_{0.75}$ by
chemical analysis.\cite{jirak} To check a possibility of such charge
ordering, we examined our powder pattern profiles in detail, but it was
not possible to detect any indication of the superlattice reflections
for the charge ordering in the powder diffraction data.  A study of a
single crystal sample is strongly desirable to elucidate the nature of a
possible 4/5 or 3/4 charge ordering in the C-type AFM phase.

As for the charge ordering for $x \ge 1/2$, an interesting charge
ordering with large periodicity was recently observed in the
La$_{1-x}$Ca$_{x}$MnO$_{3}$
system. \cite{ramirez,chen,chen2,mori2,mori3} In this system,
incommensurate superlattice peaks were observed at the wave vector
$Q=(\delta,0,0)$ with $\delta \sim 1-x$ below the charge ordering
temperature $T_{\rm CO}$ by electron diffraction.  To understand the
incommensurability, a stripe-type charge/orbital ordering was
proposed. \cite{cheong,mori2,mori3} In this model, a pair of
Mn$^{3+}$O$_{6}$ stripes are formed, and they are separated by another
stripe-shaped region of the Mn$^{4+}$O$_{6}$ octahedra.  A pair of
Mn$^{3+}$O$_{6}$ stripes are accompanied with the
$d(3x^2-r^2)/d(3y^2-r^2)$ orbital ordering and with a large lattice
contraction due to the JT effect, while the Mn$^{4+}$O$_{6}$ regions are
free from lattice distortions.  At $x = 1/2$, this incommensurate pairs
of JT stripes converges to the well-known CE-type orbital/spin ordering.
Similar incommensurate superlattice peaks were also observed in
Bi$_{1-x}$Ca$_{x}$MnO$_{3}$ single crystals with $0.74 \le x \le 0.82$,
\cite{bao} while the long-period structure with four-fold periodicity
(21 \AA) and 32-fold periodicity (170 \AA) to the orthorhombic lattice
unit were clearly observed in the $x=0.80$ sample.\cite{murakami} These
results also share many features with the paired JT stripes proposed for
the La$_{1-x}$Ca$_{x}$MnO$_{3}$ with $x \gtrsim 0.5$.

We would like to point out that this type of paired JT stripe ordering
can be easily excluded from the possible charge ordering for the present
C-type AFM Nd$_{1-x}$Sr$_{x}$MnO$_{3}$ samples from the consideration of
the lattice parameters.  For the CE-type charge ordering and associated
paired JT stripe ordering, the orthorhombic $c$ axis (in the {\it Pbnm}
notation) must be the shortest.  This is due to the fact that the
$d(3x^2-r^2)/d(3y^2-r^2)$ orbitals lie in the $ab$ plane as shown in
Fig. \ref{orbital}, and it is easily checked that this relation is
satisfied by other manganites with the CE-type ordering, for example,
Pr$_{1-x}$Ca$_{x}$MnO$_{3}$ (Refs. \onlinecite{jirak,yoshi2,lees}) as
well as La$_{1-x}$Ca$_{x}$MnO$_{3}$ with $x \geq 0.50$ (Refs.
\onlinecite{wollan,radaelli,ibarra}) from the lattice constants data in
the existing reports.

On the other hand, the $c$ axis is the longest for the C-type AFM spin
ordering in the present Nd$_{1-x}$Sr$_{x}$MnO$_{3}$, in the C-type AFM
region of Pr$_{1-x}$Ca$_{x}$MnO$_{3}$ (Ref. \onlinecite{jirak}) and in
La$_{0.2}$Ca$_{0.8}$MnO$_{3}$ (Ref. \onlinecite{wollan}).  As explained
in the previous subsection, this relation of the lattice parameters
results from the $d(3z^2-r^2)$-type orbital ordering in the C-type
structure.  Concerning the magnetic ordering of the
Bi$_{1-x}$Ca$_{x}$MnO$_{3}$ system, we are puzzled that the magnetic
ordering was reported to be of C-type.\cite{bao} Because the $c$ axis
ought to be the longest for the C-type spin order, it cannot be
compatible with the proposed JT stripe-type ordering with the shortest
$c$ axis.

Finally, we comment that the FWHM of the (004) peak of the $x=0.75$
sample at 330 K is larger than the one at about 300 K (see
Fig. \ref{width}).  This is because a finite amount of the scattering
exists between the (220) and (004) peaks.  Similar extra scattering is
observed at some other scattering angles.  Presumably, another phase
with very close length of the $a$ and $c$ axis may exist at higher
temperatures, and the remnant of the higher $T$ phase may exist at 330
K.

\section{Conclusions}

Neutron diffraction study was performed on Nd$_{1-x}$Sr$_{x}$MnO$_{3}$
powder samples with $0.49 \le x \le 0.75$ and their crystal and magnetic
structure were analyzed by the Rietveld method. A systematic change of
the crystal and magnetic structures was observed as a function of $x$.
With increasing $x$, the magnetic structure of the ground state varies
from metallic ferromagnetism to charge ordered CE-type
antiferromagnetism, then to metallic A-type antiferromagnetism, and
finally to insulating C-type antiferromagnetism. The magnetic structure
is driven by underlying Mn $e_g$ orbital ordering and resultant crystal
structure.  In the CE-type and A-type AFM states, the crystal structure
is characterized by apically compressed MnO$_{6}$ octahedra reflecting
the planar $d(x^2-y^2)$-type orbital ordering. On the other hand, in the
C-type AFM state it consists of apically elongated octahedra which is
influenced by the ordering of the rod-type $d(3z^2-r^2)$ orbitals.

The CE-type AFM state was observed only in the neighborhood of
$x=1/2$. In the $x=0.51$ sample, the CE-type AFM state and the A-type
AFM state coexisted due to the small energetic difference between the
two AFM states. In addition, the C-type AFM phase exhibits an
anisotropic broadening of Bragg peaks, which becomes clearer as $x$
increases.  This can be interpreted as a precursor of the
$d(3z^2-r^2)$-type orbital ordering at $\mbox{Mn}^{3+}:\mbox{Mn}^{4+} =
1:3$ or $1:4$.

\acknowledgements

This work was supported by a Grand-In-Aid for Scientific Research from
the Ministry of Education, Science, Sports and Culture, Japan and by the
New Energy and Industrial Technology Development Organization (NEDO) of
Japan.


\begin{references}

\bibitem[*]{kawano_ad} present address: Solid State Division, Oak Ridge
 National Laboratory, Oak Ridge, Tennessee 37831.

\bibitem[**]{kuwahara_ad} present address: Faculty of Science and
 Technology, Sophia University, Chiyoda-ku, Tokyo 102-8554, Japan.

\bibitem[***]{ohashi_ad} present address: Faculty of Engineering,
 Yamagata University, Yonezawa, Yamagata 990-8510, Japan.

\bibitem{wollan} E. O. Wollan and W. C. Koehler, Phys. Rev. {\bf 100},
 545 (1955).

\bibitem{goodenough} J. B. Goodenough, Phys. Rev. {\bf 100}, 564 (1955).

\bibitem{jirak} Z. Jir\'ak, S. Krupi\v{c}ka,
 Z. \v{S}im\v{s}a, M. Dlouh\'a, and S. Vratislav,
 J. Magn. Magn. Matt. {\bf 53}, 153 (1985).

\bibitem{cheong} S-W. Cheong and C. H. Chen, {\it Colossal
 Magnetoresistance, Charge Ordering and Related Properties of Manganese
 Oxides} ed. by C. N. R. Rao and B. Raveau (World Scientific, 1998),
 pp. 241.

\bibitem{maezono} R. Maezono, S. Ishihara, and N. Nagaosa, Phys. Rev. B
 {\bf 57}, R13993 (1998); R. Maezono S. Ishihara, and N. Nagaosa,
 Phys. Rev. B {\bf 58}, 11583 (1998).

\bibitem{mizokawa} T. Mizokawa and A. Fujimori, Phys. Rev. B {\bf 56},
 R493 (1997).

\bibitem{koshibae} W. Koshibae, Y. Kawamura, S. Ishihara, S. Okamoto,
 J. Inoue, and S. Maekawa, J. Phys. Soc. Jpn. {\bf 66}, 957 (1997).

\bibitem{kawano} H. Kawano, R. Kajimoto, H. Yoshizawa, Y. Tomioka,
 H. Kuwahara, and Y. Tokura, Phys. Rev. Lett. {\bf 78}, 4253 (1997);
 H. Kawano, R. Kajimoto, H. Yoshizawa, J.A. Fernandez-Baca, Y. Tomioka,
 H. Kuwahara, and Y. Tokura, Physica B {\bf 241-243}, 289 (1998).

\bibitem{kawano2} H. Kawano, R. Kajimoto, H. Yoshizawa, Y. Tomioka,
 H. Kuwahara, and Y. Tokura, cond-mat/9808286.

\bibitem{kuwahara} H. Kuwahara, Y. Tomioka, A. Asamitsu, Y. Moritomo,
 and Y. Tokura, Science {\bf 270}, 961 (1995).

\bibitem{kuwahara2} H. Kuwahara, T. Okuda, Y. Tomioka, A. Asamitsu, and
 Y. Tokura, Mat. Res. Oc. Symp. Proc. {\bf 494}, 83 (1998).

\bibitem{kuwahara3} H. Kuwahara, T. Okuda, Y. Tomioka, A. Asamitsu, and
 Y. Tokura, unpublished, and private communications.

\bibitem{kuwahara4} H. Kuwahara, Y. Moritomo, Y. Tomioka, A. Asamitsu,
 M. Kasai, R. Kumai, and Y. Tokura, Phys. Rev. B {\bf 56}, 9386 (1997).

\bibitem{izumi} F. Izumi, {\it The Rietveld Method,} ed. by R. A. Young,
 (Oxford University Press, Oxford, 1993), Chap. 13; Y.-I. Kim and
 F. Izumi, J. Ceram. Soc. Jpn., {\bf 102}, 401 (1994).

\bibitem{O_prime} Although the relation of the $a$ and $b$ parameters is
 reverse to the difinition in Ref. \onlinecite{jirak}, we shall call
 this structure as O$^{\prime}$ structure in this paper.

\bibitem{jirak2} Z. Jir\'ak, E. Pollert, A. F. Andersen, J.-C. Grenier,
 and P. Hagenmuller, Eur. J. Solid State Inorg. Chem. {\bf 27}, 421
 (1990).

\bibitem{caignaert} V. Caignaert, F. Millange, M. Hervieu, E. Suard, and
 B. Raveau, Solid State Comm. {\bf 99}, 173 (1996).

\bibitem{wolfman} J. Wolfman, A. Maignan, Ch. Simon, and B. Raveau,
 J. Magn. Magn. Matt. {\bf 159}, L299 (1996).

\bibitem{sundaresan} A. Sundaresan, P. L. Paulose, R. Mallik, and
 E. V. Sampathkumaran, Phys. Rev. B {\bf 57}, 2690 (1998).

\bibitem{glazer} A. M. Glazer, Acta Crystallogr. A {\bf 31}, 756 (1975).

\bibitem{yoshi2} H. Yoshizawa, H. Kawano, Y. Tomioka, and Y. Tokura,
 Phys. Rev. B {\bf 52}, R13145 (1995); H. Yoshizawa, H. Kawano,
 Y. Tomioka, and Y. Tokura, J. Phys. Soc. Jpn. {\bf 65}, 1043 (1996).

\bibitem{kaji98} R. Kajimoto, T. Kakeshita, Y. Oohara, H. Yoshizawa,
 Y. Tomioka, and Y. Tokura, Phys. Rev. B {\bf 58}, R11837 (1998).

\bibitem{cox} D. E. Cox, P. G. Radaelli, M. Marezio, and S-W. Cheong,
 Phys. Rev. B {\bf 57}, 3305 (1998).

\bibitem{mor98a} An {\it metallic} A-type AFM state is also reported by
 T. Akimoto, Y. Maruyama, Y. Moritomo, A. Nakamura, K. Hirota,
 K. Ohoyama, and M. Ohashi, Phys. Rev. B {\bf 57}, R5594 (1998).

\bibitem{mor98}Y. Moritomo, T. Akimoto, A. Nakamura, K. Ohoyama, and
 M. Ohashi, Phys. Rev. B {\bf 58}, 5544 (1998).

\bibitem{com}Close inspection of the crystal structures of the CE-type
 samples and the A-type $x=0.55$ sample leads one to notice that there
 is no significant difference of the structural parameters except a
 smooth variation as a function of $x$.

\bibitem{yoshi} H. Yoshizawa, H. Kawano, J. A. Fernandez-Baca,
 H. Kuwahara, and Y. Tokura, Phys. Rev. B {\bf 58}, R571 (1998).

\bibitem{cant_or_not} As far as the diffraction condition is concerend,
 it is possible that the canting angle is 0$^{\circ}$. However, from the
 value of the magnetic moments listed in Table \ref{moment_table},
 0$^{\circ}$ canting angle implies the abnormal moment distribution of
 $\sim 4 \mu_{\rm B}$ and $\sim 0.9 \mu_{\rm B}$, and it is very
 unlikely.

\bibitem{stern}B. J. Sternlieb, J. P. Hill, U. C. Wildgruber, G. M.
Luke, B. Nachumi, Y. Moritomo, and Y. Tokura Phys. Rev. Lett. {\bf 76},
2169 (1996). 

\bibitem{kubota} M. Kubota, H. Yoshizawa, H. Fujioka, K. Hirota,
 Y. Moritomo, and Y. Endoh, cond-mat/9811192.

\bibitem{com2}We noticed that the profiles of several peaks are
broad and asymmetric.

\bibitem{ramirez} A. P. Ramirez, P. Schiffer, S-W. Cheong, C. H. Chen,
 W. Bao, T. T. M. Palstra, P. L. Gammel, D. J. Bishop, and B. Zegarski,
 Phys. Rev. Lett., {\bf 76}, 3188 (1996).

\bibitem{chen} C. H. Chen and S.-W. Cheong, Phys. Rev. Lett. {\bf 76},
 4042 (1996).

\bibitem{chen2} C. H. Chen, S-W. Cheong, and H. Y. Hwang,
 J. Appl. Phys. {\bf 81}, 4326 (1997).

 \bibitem{mori2} S. Mori, C. H. Chen, and S-W. Cheong, Nature {\bf 392},
 473 (1998).

 \bibitem{mori3} S. Mori, C. H. Chen, and S-W. Cheong,
 Phys. Rev. Lett. {\bf 81}, 3972  (1998).

\bibitem{bao} W. Bao, J. D. Axe, C. H. Chen, and S-W. Cheong,
 Phys. Rev. Lett. {\bf 78}, 543 (1997).

\bibitem{murakami} Y. Murakami, D. Shindo, H. Chiba, M. Kikuchi, and
 Y. Shono, Phys. Rev. B {\bf 55}, 15043 (1997).

\bibitem{lees} M. R. Lees, J. Barratt, G. Balakrishnan, D. McK. Paul,
 and C. Ritter, Phys. Rev. B {\bf 58}, 8694 (1998).

\bibitem{radaelli} P. G. Radaelli, D. E. Cox, M. Marezio, S-W. Cheong,
 Phys. Rev. B {\bf 55}, 3015 (1997);  P.G. Radaelli, D.E. Cox,
 L. Capogna, S-W. Cheong, M. Marezio, cond-mat/9812366.

\bibitem{ibarra} M. R. Ibarra, J. M. De Teresa, J. Blasco,
 P. A. Algarabel, C. Marquina, J. Garcia, J. Stankiewicz, and C. Ritter,
 Phys. Rev. B {\bf 56}, 8252 (1997).

\end{references}
\end{document}